\documentclass[letterpaper]{article} 
\usepackage{aaai23}
\usepackage{times}  
\usepackage{helvet}  
\usepackage{courier}  
\usepackage[hyphens]{url}  
\usepackage{graphicx} 
\usepackage{natbib}  
\usepackage{caption} 
\usepackage{color}
\usepackage{algorithm}
\usepackage{algorithmic}

\usepackage{tabularx}
\usepackage{multirow}
\usepackage{booktabs}

\usepackage{newfloat}
\usepackage{listings}
\DeclareCaptionStyle{ruled}{labelfont=normalfont,labelsep=colon,strut=off} 
\floatstyle{ruled}
\newfloat{listing}{tb}{lst}{}
\floatname{listing}{Listing}
\usepackage{amsmath}
\usepackage{amsthm,amssymb}

\title{UniSyn: An End-to-End Unified Model for Text-to-Speech and Singing Voice Synthesis}
\author {
    Yi Lei\textsuperscript{\rm 1},
    Shan Yang\textsuperscript{\rm 2},
    Xinsheng Wang\textsuperscript{\rm 1},
    Qicong Xie\textsuperscript{\rm 1},
    Jixun Yao\textsuperscript{\rm 1},
    Lei Xie\textsuperscript{\rm 1}\thanks{Corresponding author.},
    Dan Su\textsuperscript{\rm 2}
}
\affiliations {
    \textsuperscript{\rm 1} Audio, Speech and Language Processing Group (ASLP@NPU), School of Computer Science,\\
  Northwestern Polytechnical University, Xi’an, China\\
    \textsuperscript{\rm 2} Tencent AI Lab, China\\
    leiyi@npu-aslp.org,
    \{xieqicong, yaojx\}@mail.nwpu.edu.cn,
    lxie@nwpu.edu.cn,\\ 
    w.xinshawn@gmail.com,
    \{shaanyang, dansu\}@tencent.com
}

\usepackage{bibentry}

\begin{document}

\maketitle

\begin{abstract}

Text-to-speech (TTS) and singing voice synthesis (SVS) aim at generating high-quality speaking and singing voice according to textual input and music scores, respectively. Unifying TTS and SVS into a single system is crucial to the applications requiring both of them. Existing methods usually suffer from some limitations, which rely on either both singing and speaking data from the same person or cascaded models of multiple tasks. To address these problems, a simplified elegant framework for TTS and SVS, named \textit{UniSyn}, is proposed in this paper. It is an end-to-end unified model that can make a voice speak and sing with only singing or speaking data from this person. To be specific, a multi-conditional variational autoencoder (MC-VAE), which constructs two independent latent sub-spaces with the speaker- and style-related (i.e. speak or sing) conditions for flexible control, is proposed in UniSyn. Moreover, supervised guided-VAE and timbre perturbation with the Wasserstein distance constraint are leveraged to further disentangle the speaker timbre and style. Experiments conducted on two speakers and two singers demonstrate that UniSyn can generate natural speaking and singing voice without corresponding training data. The proposed approach outperforms the state-of-the-art end-to-end voice generation work, which proves the effectiveness and advantages of UniSyn.

\end{abstract}

\section{Introduction}
Recent advances in text-to-speech (TTS)~\cite{ren2019fastspeech, wang2017tacotron,li2019neural,chen2021adaspeech,liu2021vara,jeong2021diff,weiss2021wave} and singing voice synthesis (SVS)~\cite{nishimura2016singing,kim2018korean,blaauw2020sequence,liu2022diffsinger,gu2021bytesing,lu2020xiaoicesing,ren2020deepsinger} have greatly improved the quality and naturalness of the generated speech and singing voices, which results in the widespread applications of TTS and SVS.
The increasing requirements in real-world applications necessitate the ability for versatile voices that can not only speak but also sing. Intuitively, training such a system needs high-quality speech and singing voice from the same speaker. However, obtaining such a corpus is more difficult than the corpus consisting of speaking and singing voice recorded by different persons, e.g., it is non-trivial to build a singing corpus for a person who is not good at singing. While the multi-task cascaded system consisting of TTS and singing voice conversion (SVC)~\cite{Zhang2020,guo2022improving,liu2021fastsvc} models can achieve this goal, it leads to a complex pipeline and inflexible generation depending on reference audio. 
 
Both TTS and SVS share a similar pipeline that generates vocal voices from abstract information, i.e., textual input or musical scores, thus making it intuitively reasonable to realize the SVS and TTS tasks with a unified model that enables a target speaker timbre to speak and sing with the absence of speech or singing data. However, several challenges will rise to unify TTS and SVS into one model. First, as the important input information, musical scores in the SVS task are quite different from the textual input in the TTS task. Second, although both speaking and singing voices are generated from the same vocal articulation system, the acoustic outcomes have substantial differences in both timbre and prosodic aspects. Third, the speaker timbre and style (i.e. speak and sing) of source audio are heavily entangled. When only one style from each person is accessible in the training phase, this entanglement tends to result in the timbre leakage issue~\cite{lee2020disentangling,xue2021learn2sing}, which means the synthesized singing voice sounds like the timbre that provides the singing training data rather than the target speaker that only has speech training data and vice versa. Due to these challenges, previous efforts to build the unified TTS and SVS model did not obey the typical pipeline but utilized the explicit features (e.g. pitch contour and rhythm) extracted from the waveform as input to generate the desired speaking or singing voice for the target speaker timbre~\cite{valle2020mellotron,Zhang2020,xue2021learn2sing}. However, the reference signals are always essential for providing explicit features at inference time. Therefore, the generated singing voice relies on the reference and cannot synthesize arbitrary songs, which reduces the flexibility in practical applications.

In this paper, we develop \textit{UniSyn}, which is an end-to-end unified system for both speech and singing voice synthesis with only speaking or singing data from each person. UniSyn consists of three components: 1) a variational autoencoder to encode the waveform into a latent space, 2) a prior model to estimate the latent distribution from the input linguistic representations, and 3) a wave decoder to generate the waveform from sampled latent variables.

To overcome the challenges of unified modeling, we propose unified linguistic representations for the text content of TTS and the music score of SVS. To facilitate speaking/singing style transfer cross speakers in the unified model, we propose a multi-conditional variational autoencoder (MC-VAE) to lead the learned latent space more interpretable by being divided into two independent sub-spaces. One sub-space is for speaker conditioned on speaker identity, and the other is for the remaining information in the voice conditioned on other input representations except for speaker identity~\cite{ding2020guided}.
To better decouple the speaker timbre and style (speaking/singing), we adopt the speaker identity and pitch contour as supervision to conduct disentanglement on the latent variables. Furthermore, the timbre perturbation strategy~\cite{choi2021neural} with the Wasserstein distance constraint is also utilized to auxiliarily learn the remaining sub-space for improving the robustness of the disentangled speaker timbre. Note that different from the well-settled two-stage voice generation system that adopts an independently trained acoustic model and neural vocoder to output waveform, UniSyn is an \textit{end-to-end} trained neural generation model that directly produces waveform.


The main contributions of our work are summarized as follows:
\begin{itemize}
    \item We propose UniSyn, an end-to-end unified system for text-to-speech and singing voice synthesis. It leverages MC-VAE to learn an interpretable latent space for flexible control and utilizes guided-VAE and formant perturbation to conduct better disentanglement. 
    \item We conduct experiments on both TTS and SVS. The results demonstrate that UniSyn significantly outperforms the state-of-the-art end-to-end generation model on SVS for speakers without singing training data, as well as TTS for singers without speech training data.
    \item To the best of our knowledge, this is the first work on speech and singing voice synthesis unified system in the end-to-end way, which can not only teach the speaker without singing data to sing but also teach the singer without speech data to speak. 
\end{itemize}

\section{Background}

\subsection{Unified Speech \& Singing Generation}
Singing voice conversion (SVC) is a common solution for making a voice sing and speak without his/her singing or speaking data. 
Recent works~\cite{saito2018non,Polyak2020,guo2022improving,li2021ppg} use the Phonetic PosteriorGrams (PPGs) extracted from audio to represent the linguistic content and eliminate the speaker characteristics. With PPGs and F0 modeling, these methods can convert the timbre of the singing voice while preserving the linguistic content. A framework~\cite{Zhang2020} is designed by unifying the features of speech and singing synthesis and hiring a speaker verification-based speaker encoder to disentangle the speaker timbre from the audio signal. The extracted speaker embedding is conditioned on the generation model to synthesize speech and singing voice in the target timbre. For converting speaker timbre of the singing voice, VAE is employed in ~\cite{luo2020singing} to disentangle speaker timbre from audio in the latent space. 
The literature on the unified model for TTS and SVS without relying on speech or singing training data usually utilizes the explicit features extracted from the audio signal as the input representations. Mellotron~\cite{valle2020mellotron} is proposed for both speech and singing voice synthesis by explicitly conditioning on the melodic information such as pitch and rhythm. However, reference audio is necessary for providing explicit features of speech or singing at inference time, which is inflexible for synthesizing arbitrary content. Instead of reference audio, approaches~\citet{xue2022learn2sing,liu2021vibrato} are proposed to utilize the music score to produce singing voice conditioned on the speaker identity, whereas they cannot generate speech since the non-universal input. In this paper, we unify the textual content of speech and music score of singing and construct a unified model for TTS and SVS without leveraging any reference audio when inference.

\subsection{VAE-based Speech Generation}
VAE is a generation model that extracts a latent distribution of the data and reconstructs the data from the sampled latent variables. The VAE family has been successfully applied to TTS and SVS.
Recently, the conditional-VAE-based voice synthesis model, VITS~\cite{kim2021conditional}, has brought success to high-quality speech generation in a parallel end-to-end TTS framework. VITS adopts a variational autoencoder to approximate the latent distribution of waveform $x$ and a text encoder followed by a stochastic duration predictor to encode the text sequence for producing a condition $c$. Then a normalizing flow-based decoder is used to generate waveform from the sampled latent variables $z$. VITS can be expressed in the CVAE process by optimizing the evidence lower bound (ELBO) of the interactable marginal log-likelihood of waveform $\log p\left( x|c \right)$:
\begin{equation}
\label{eq:cvae}
\begin{aligned}
  & \log p \left( x|c \right) \geq  \mathbb{E}_{q \left(z|x \right)} \left [ \log p \left(x|z\right)  - \log \frac{q \left(z|x \right)}{p \left(z|c \right)} \right],
\end{aligned}
\end{equation}
where $p \left( x|z \right)$ is the likelihood function that generate waveform $x$ given the latent variables $z$, $q \left(z|x \right)$ is the posterior distribution approximated by the posterior encoder, $p \left(z|c \right)$ is the prior distribution of latent variables $z$ given the condition $c$.
Because of the success of VITS, its framework has been extended to SVS. 
By replacing the Monotonic Alignment Search (MAS), ~\citet{zhang2022visinger} introduce a length regulator with a duration predictor~\cite{ren2019fastspeech} and a frame prior network to predict a frame-level distribution for natural singing voice synthesis in an end-to-end way. However, these works leverage flow-decoder to disentangle the speaker information, which may reduce the robustness of the generation system and lack of interpretability of the latent space.
In this paper, we improve the CVAE-based end-to-end synthesis framework by conducting disentanglement in the latent space for a more robust generation.



\begin{figure*}[ht]
        \centering
        \includegraphics[width=1.0\linewidth]{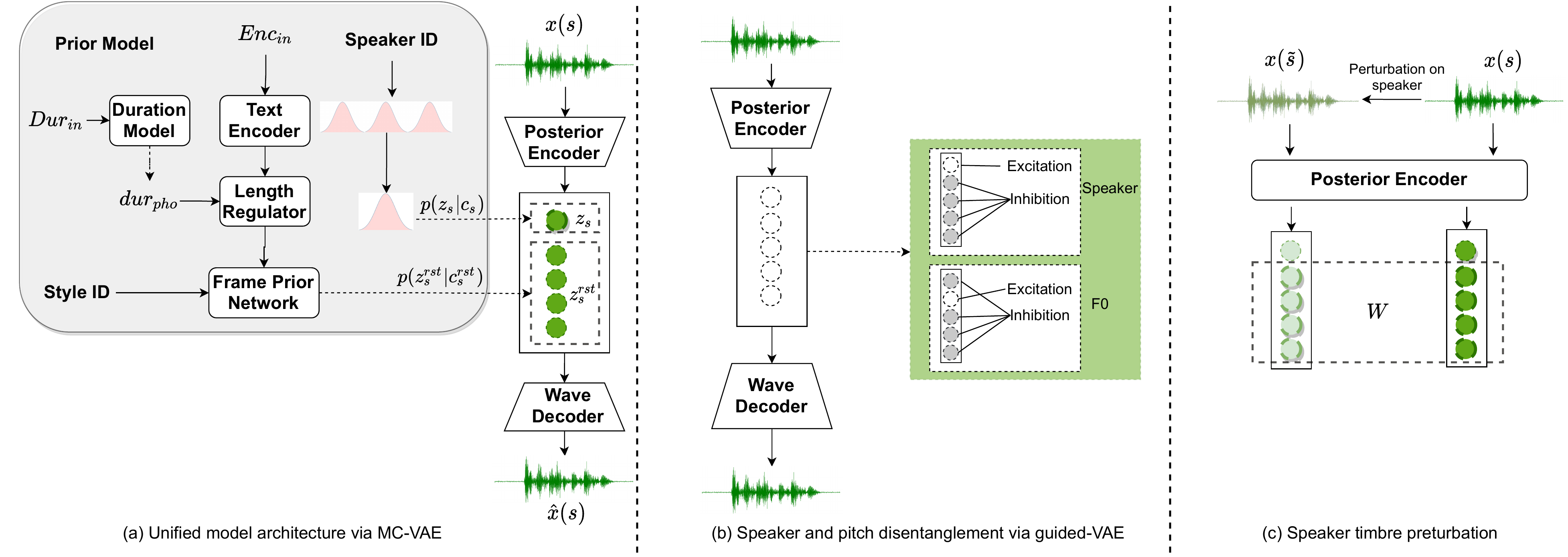}
        \caption{The architecture of UniSyn. In subfigure (a), $Enc_{in}$ is the unified input label for text encoder; $Dur_{in}$ is the unified input for duration prediction; \textit{Speaker ID} is the speaker identity; \textit{style ID} is the style identity for speaking or singing; $x(s)$ is the waveform of the speaker $s$; $\hat{x}(s)$ is the reconstructed waveform; the dotted lines are the KL divergence. In subfigure (b), $Excitation$ and $Inhibition$ are the excitation function and inhibition function in the guided-VAE respectively. In subfigure (c), $x(\widetilde{s})$ is the speaker perturbed waveform; $W$ is the Wasserstein distance loss between the latent distributions.
        }
         \vspace{-10pt}
        \label{fig:system}
\end{figure*}

\section{UniSyn}
\label{Model architecture}
We will introduce the proposed unified model for cross-speaker speech and singing voice generation as illustrated in Figure~\ref{fig:system}, including 1) the unified textual information modeling for TTS and SVS, 2) the proposed MC-VAE for flexible multi-speaker voice generation with multiple conditions, 3) the supervised guided-VAE to disentangle speaker timbre and pitch contour from the latent representations of audio signals, and 4) the timbre perturbation strategy with Wasserstein distance constraint to further enhance the disentanglement ability on the timbre of speech. 

\subsection{Unified Textual Features for TTS and SVS}
\label{sc:label}

Previous work~\cite{valle2020mellotron} has shown that both speaking and singing voice could be simultaneously decomposed into explicit acoustic features like rhythm, pitch, speaking style, and text content. To achieve the unified voice generation, we design a unified textual representation for speaking and singing. 
In TTS, the textual information often can be factorized as \textit{phoneme}, \textit{tone} (for tonal language), and \textit{phoneme duration}. While in SVS, the music score usually includes \textit{phoneme}, \textit{note pitchID}, \textit{phoneme duration}, and \textit{note duration}. Aiming at building a unified voice synthesis model, it's necessary to factorize the common and different aspects between the text and music score inputs. 

For the common part, 1) the linguistic content can be defined as the \textit{phoneme} attribute for both speech and singing, which is marked as $pho$ in our model; 2) the melody variations of speaking and singing voice are defined as an attribute named $tp$ merged by the \textit{note pitchID} of the music score and the \textit{tone} of input text, since both they determine the melody acoustic feature. 
Noted that for non-tonal languages, there is no explicit input feature representing speech melody, so the \textit{note pitchID} should be treated as a different feature in unified textual representations. 

With the basic pronunciation skeleton, the phoneme duration $dur_{pho}$ defines the rhythm of both speech and singing. But unlike speech, the duration of a phoneme in singing voice is bound by the music notes $dur_{note}$ in the music scores. To form a unified representation, we set a placeholder of $dur_{note}$ for speaking voice.
To unify speaking and singing prosody rhythm, a shared duration predictor is leveraged for modeling the phoneme duration of both speech and singing voices. Intuitively, $dur_{pho}$ is strong affected by $pho$, $dur_{note}$, and the style tag. Besides, we observe that in Mandarin Chinese, the vowel always has a longer duration than the consonant in the same syllable, especially in singing voices. Therefore, we introduce a \textit{relative position} attribute $pos$ for a more accurate estimation of phoneme duration in the duration model, which is obtained based on the total number of phonemes and the rank of the current phoneme in the current music note or speaking syllable.

As a summary, the attribute set $Enc_{in} = \{ pho, tp \}$ is involved to model the linguistic information in the text encoder, while the duration predictor takes the $Dur_{in} = \{pho, dur_{note}, pos, style \}$ to predict the phoneme duration for both speech and singing, which is then applied in the Length Regulator to match the time resolution of acoustic latent features.

\subsection{MC-VAE for Voice Generation}
\label{sc:mcvae}
For building an end-to-end generation system, we follow the CVAE-based skeleton in VITS~\cite{kim2021conditional} to produce waveform from the latent variables. In this work, we separate the latent space into two sub-spaces to make the latent variables learned from VAE more interpretable. 
Ideally, we tend to decompose the audio signal into two independent attributes, named speaker timbre and remaining attributes. To achieve this goal, we propose the multi-conditional variational autoencoder (MC-VAE). 


Given input waveform $x$, we can learn its latent distribution $q(z|x)$ through conventional VAE encoder. In MC-VAE, we try to divide the latent $z$ into two independent spaces $z = (z_s, z_s^{rst})$, where $z_s$ only defines a speaker subspace and $z_s^{rst})$ denotes the remaining information in waveform such as linguistic content and style. To make the $z_s$ and $z_s^{rst}$ independent, we utilize two independent conditions $c=(c_s, c_s^{rst})$ to restrict the sub-spaces, where $c_s$ is the categorical speaker label $speaker ID$ for modeling $z_s$ and $c_s^{rst}$ is the remaining information generated from the frame-level speaker-independent prior network for modeling $z_s^{rst}$.


We define the prior distribution of speaker follows the conditional Gaussian distribution $p(z_s|c_s) = N(\mu_{c_{s}}, \sigma_{c_{s}})$, where $\mu_{c_{s}}$ and $\sigma_{c_{s}}$ are the prior mean and variance and only determined by the speaker ids. Meanwhile, the remaining information $z_s^{rst}$ of audio can be approximated from the unified textual features and the style embedding, which we denote as $c_s^{rst}$. With the independent conditions $c=(c_s, c_s^{rst})$ and independent latent variables $z=(z_s, z_s^{rst})$, by the conditionally independent property of joint probability distributions, the variational lower bound of MC-VAE in our model can be written as:

\vspace{-3pt}
\begin{equation}
\label{eq:mcvae}
\begin{aligned}
 & \mathbf{ELBO} \left( p, q; x, c \right) 
 \\ & = \mathbb{E}_{q \left(z|x \right)} \left [ \log p \left(x|z\right)  - \log \frac{q \left(z|x \right)}{p \left(z|c \right)} \right]
  \\ & = \mathbb{E}_{q \left(z|x \right)} \left [ \log p \left(x|z\right)\right] - \mathbb{E}_{q \left(z_s, z_r|x \right)} \left [\log \frac{q \left(z_s, z_s^{rst}|x \right)}{p \left(z_s, z_s^{rst}|c \right)} \right]
   \\ & = \mathbb{E}_{q \left(z|x \right)} \left [ \log p \left(x|z\right)\right] 
   \\& \quad - \mathbb{E}_{q \left(z_s|x \right) q \left(z_s^{rst}|x \right)} \left [\log \frac{q \left(z_s|x \right)  q \left(z_s^{rst}|x \right) }{p \left(z_s|c_s \right)  p \left(z_s^{rst}|c_s^{rst} \right)} \right]
 \\ & = \mathbb{E}_{q \left(z|x \right)} \left [ \log p \left(x|z\right)\right]  
 \\ &  \quad - KL\left( q \left(z_s|x\right) || p \left(z_s |c_s \right) \right)  
 \\ & \quad - KL\left( q \left(z_s^{rst}|x\right) || p \left(z_s^{rst} |c_s^{rst} \right) \right),
\end{aligned}
\end{equation}
where the first term is the reconstruction loss ${\cal L}_{rec}$. The second and third terms are KL divergences, marked as ${\cal L}_{kl}^{z_s}$ and ${\cal L}_{kl}^{z_s^{rst}}$ respectively.
We use the $L_1$ loss between the mel-spectrogram $mel_x$ from ground-truth waveform $x$ and $mel_{\hat{x}}$ from the predicted waveform $\hat{x}$ in frequency domain to reconstruct $x$:
\begin{equation}
   {\cal L}_{rec} = \left\| {mel_{x} - { mel_{\hat x}}} \right\|_1.
\end{equation} 

${\cal L}_{kl}^{z_s}$ is a reverse-KL divergence to estimate the true distribution $p(z_s|c_s)\sim N(\mu_{c_{s}}, \sigma_{c_{s}})$ with our approximate distribution $q(z_s|x)$. The reverse-KL divergence makes objective \textit{mode seeking}~\cite{vaswani2022general,mei2019principled} that helps the approximate distribution $q(z_s|x)$ to find an accurate speaker with high probability and mimic it exactly. On the contrary, ${\cal L}_{kl}^{z_s^{rst}}$ is a forward-KL divergence to estimate the posterior distribution $q(z_s^{rst}|x)$ with the approximate prior distribution $p(z_s^{rst}|c_s^{rst})$, which is a \textit{mode covering}~\cite{agarwal2019learning} instance that is suitable for learning the timbre-independent remaining information.


Since the timbre of speech is almost invariant over time, the variance of the speaker distribution should be relatively smaller. We directly treat the speaker id $c_s$ as $\mu_{c_{s}}$ and set $\sigma_{c_{s}}$ is equal to 0.01, where the smaller variance is also benefit to distinguish speaker with the Gaussian distribution.

\subsection{Supervised Guided-VAE for Disentanglement}
\label{sc:gvae}
To disentangle the desired attribute, such as speaker timbre, from the latent variables $z$, we utilize a supervised guided-VAE (GVAE)~\cite{ding2020guided}, as shown in Figure~\ref{fig:system}(b).
For an attribute $f$ that needs to be disentangled from waveform, GVAE defines the latent variable from the posterior encoder as $z=(z_f, z_f^{rst})$, where $z_f$ is a scalar variable deciding this attribute and $z_f^{rst}$ represents the remaining latent variables. The objective of GVAE contains an adversarial excitation and inhibition method as:
\begin{equation}
\begin{aligned}
    {\cal L}_{excitation}^f = {L}_{pred} (\text{Pred}_{f}(z_f), f_x)
\end{aligned}
\end{equation}
\begin{equation}
\begin{aligned}
    {\cal L}_{inhibition}^f = {L}_{pred} (\text{Pred}_{f}(z_f^{rst}), f_x)
\end{aligned}
\end{equation}
\begin{equation}
\label{eq:gvae}
\begin{aligned}
   {\cal L}_{gvae}^f = {\cal L}_{excitation}^f + 1 \bigg/ {\cal L}_{inhibition}^f,
\end{aligned}
\end{equation}
where $f_x$ is the ground-truth value of the attribute $f$ of waveform $x$, $\text{Pred}_{f}$ refers to the network to predict the attribute value from latent $z$, and ${L}_{pred}$ denotes the loss function to optimize prediction results. By minimizing the Eq.~\ref{eq:gvae}, the excitation process encourages $z_f$ containing the attribute information $f$, and the inhibition process can be treated as an adversarial term that makes $z_f^{rst}$ as uninformative to $f$ as possible.

In Unisyn, we utilize the GVAE to disentangle both speaker timbre and pitch information, since the pitch contour is critical to singing voice generation for the speaker without singing data. Specifically, the latent $z$ can be presented as $z=(z_s, z_s^{rst})$ and $z=(z_p, z_p^{rst})$ respectively.
For speaker disentanglement, we optimize the shared latent variable $z=(z_s, z_s^{rst})$ from MC-VAE, where the \textit{cross entropy} is treated as ${L}_{pred}$ to predict the speaker identity. As for pitch disentanglement, $z=(z_p, z_p^{rst})$ is optimized to make sure the $z_p$ only contains pitch-related information, where the \textit{MSE} loss is used to predict pitch values.

\subsection{Speaker Timbre Perturbation}
\label{sc:pert}
With the above MC-VAE and GVAE, the timbre and pitch contour can be disentangled from the waveform, and the vocal timbre also can be flexibly controlled by a manual label or sampling from reference audio. 
To further improve the robustness of generating speaking or singing voice with a target timbre without corresponding training data, we further conduct information perturbation~\cite{choi2021neural} on formant of waveform to obtain speaker-independent augmented training data.

In detail, we utilize a formant shifting function $fs$ to distort the timbre of the audio at a random range, since the formant is highly related to vocal timbre. Given the waveform $x(s)$ from speaker $s$, we apply the function $fs$ on it during each training step to obtain $x(\Tilde{s})$, where only formant is randomly shifted and other information is preserved. In this way, we assume the latent variables $z_s^{rst}$ and $z_{\Tilde{s}}^{rst}$ in MC-VAE should follow the same distribution. We then utilize a Wasserstein distance~\cite{zhao2018adversarially} constraint between $z_s^{rst}$ and $z_{\Tilde{s}}^{rst}$ to encourage them to learn the speaker-independent information:
\begin{equation}
\label{eq:wloss}
   {\cal L}_{pert} =  W\left( q\left(z_s^{rst}|x(s)\right) || q \left( z_{\Tilde{s}}^{rst} |x(\Tilde{s}) \right) \right),
\end{equation}
where ${W}$ represents the Wasserstein distance between two distributions. We utilize the Wasserstein distance since it has been shown good performance on the text sequences~\cite{zhao2018adversarially}.

\subsection{Training}
\label{sc:adv} 
To further improve the quality of generated audio, we also apply the adversarial training strategy during training~\cite{kong2020hifi,kim2021conditional}. Following the state-of-the-art vocoder HiFi-GAN~\cite{kim2021conditional}, we utilize several discriminators, including a Multi-Period Discriminator (MPD) and a Multi-Scale Discriminator (MSD), to discriminate the real and fake waveform on different scales. The feature-mapping loss $\mathcal{L}_{fm}$ is also applied to the generator to constrain its output in each hidden layer of the discriminators~\cite{kong2020hifi}.

Combining the above MC-VAE, GVAE, speaker perturbation and adversarial training, we optimize the model by minimizing the loss function as 
\begin{equation}
\label{eq:totalloss}
\begin{aligned}
   {\cal L}_{G} = & \alpha{\cal L}_{rec} + \beta{\cal L}_{kl}^{z_s} + \gamma{\cal L}_{kl}^{z_s^{rst}} + \lambda{\cal L}_{gvae}
   \\ &  + \mu{\cal L}_{pert} + \eta\mathcal{L}_{fm} + \theta\mathcal{L}_{adv}(G) + \phi{\cal L}_{dur},
\end{aligned}
\end{equation}
where ${\cal L}_{gvae}$ is the sum of ${\cal L}_{gvae}^s$ and ${\cal L}_{gvae}^p$, ${\cal L}_{dur}$ is the $L_1$ loss between the predicted duration and the ground-truth duration in the log domain, and $\mathcal{L}_{adv}(G)$ is the adversarial loss for the generator. The hyper-parameters $\alpha$, $\beta$, $\gamma$, $\lambda$, $\mu$, $\eta$, $\theta$, and $\phi$ are the weights of the losses. We set $\alpha=60$, $\beta=12$, $\gamma=1.5$, $\lambda=10$, $\mu=0.02$, $\eta=2$, $\theta=2$, and $\phi=1.5$ in our model empirically.

\subsection{Inference}
\label{sc:inf} 
\begin{figure}[!htb]
        \centering
        \includegraphics[width=1.0\linewidth]{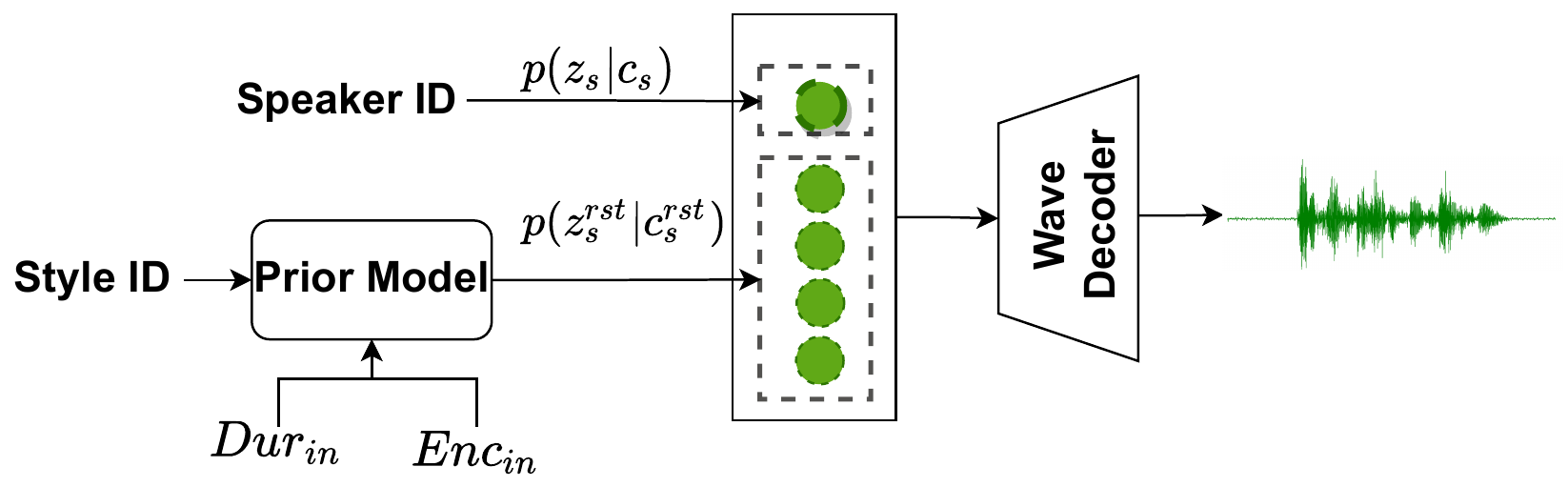}
        \caption{The inference procedure of UniSyn.}
        \label{fig:inference}
\end{figure}

The inference procedure of UniSyn is shown in Figure~\ref{fig:inference}, which supports both TTS and SVS. The speaker-related latent variable $z_s$ is sampled from the speaker prior distribution conditioned on speaker ID, and the latent $z_s^{rst}$ is encoded from the prior model with the unified textual features. With $z=(z_s, z_s^{rst})$, the wave decoder can generate corresponding speech and singing voice of the target speaker. 



\subsection{Model architecture}
\label{sc:arch}
\subsubsection{Prior Model.}
The prior model aims at providing the speaker and remaining prior distributions.
It consists of several components: 1) a \textbf{text encoder} followed by a length regulator to encode the unified textual inputs into frame-level representations, which contains 6 Feed-Forward Transformer blocks~\cite{ren2019fastspeech}; 2) a \textbf{duration predictor} including 3 convolution layers with dropout to provide each phoneme duration during inference; 3) a \textbf{frame prior network}, including 6 Transformer blocks, to produce the frame-level prior latent variables $z_s^{rst}$; 4) a speaker prior procedure to produce the $p(z_s|c_s)$. 

\subsubsection{Posterior Encoder.}
The posterior encoder contains a linear-spectrogram extractor, several WaveNet~\cite{shen2018natural} residual blocks, and a linear projection layer, aiming at extracting the mean and variance of the posterior distribution $q(z|x)$ from waveform. 


\subsubsection{Wave Decoder.}
Given the sampled $z \sim p(z|x)$ with re-parametrization trick, the wave decoder tends to reconstruct $x$ from $z$, where $z$ is sliced to fixed length for more efficient training. The wave decoder consists of a stack of transposed convolutions with the multireceptive field fusion module (MRF)~\cite{kong2020hifi} to match the resolution of audio samples.









\section{Experimental Setup}

\subsection{Dataset}
To evaluate the performance of UniSyn, we conduct experiments with a mixture of Mandarin speech and singing corpora. The singing corpora contain data from two female singers: 1) the Opencpop~\footnote{\url{https://wenet.org.cn/opencpop/}}~\cite{wang2022opencpop}, an open-source singing corpus with 100 pop songs recorded by a professional singer, which we denote as \textit{Singer-1}; and 2) an internal singing corpus with 100 songs from another female singer, denoted as \textit{Singer-2}. Both of the singing datasets have about 5-hour audio, and we split them into sentence pieces for training.
The speech corpora consist of two speakers: 1) an open-source Mandarin TTS dataset~\footnote{\url{https://www.data-baker.com/open_source.html}} recorded from a female speaker, which contains about 10-hour neutral speech, denoted as \textit{Speaker-1}; and 2) an internal Mandarin corpus recorded from another female speaker, totally about 5 hours, which we denote as \textit{Speaker-2}. 

To balance the amount of data for speech and singing, we randomly select about 2 hours of audio from each speaker for training. For validation and evaluation, 100 utterances from the rest data and two preserved songs from each singing corpus are involved. We down-sample all the speech and singing audios into 24k Hz, and set the frame size and hop size to 1200 and 300 respectively when extracting optional auxiliary acoustic features like pitch and spectrogram. The auxiliary pitch contour is extracted through WORLD~\cite{morise2016world}, and the implementation of formant shifting follows the NANSY~\cite{choi2021neural} model using Praat~\cite{boersma2001speak}. The phoneme duration is obtained through an HMM-based force alignment model~\cite{sjolander2003hmm}.


\subsection{Benchmark Systems}
We choose the state-of-the-art end-to-end speech synthesis framework VITS for comparison since VITS can conduct the cross-speaker generation by the flow-decoder to disentangle the speaker timbre. Because there is rare research to make a singer speak, we use Learn2Sing 2.0~\cite{xue2022learn2sing} to conduct a comparison on the SVS task additionally.
We separately measure the performance of TTS and SVS for the unified model. We conduct experiments on the following systems: 1) GT, the ground-truth recordings; 2) $\text{VITS}_{\text{TTS}}$, where we only train the system on the speaking data as the top-line of the unified model; 3) $\text{VITS}_{\text{SVS}}$, which is only trained with the singing data as the top-line of VITS for SVS; 4) $\text{VITS}_{\text{unify}}$, the unified model constructed on VITS using the flow-decoder for generating speaking and singing voice; 5) $\text{UniSyn}_{\text{TTS}}$, the proposed system trained on the speech data as the top-line of the proposed model; 
6) $\text{UniSyn}_{\text{SVS}}$, which is only trained with the singing data as the top-line of UniSyn for SVS; 7) $\text{UniSyn}_{\text{unify}}$, the proposed model for both TTS and SVS; 8) Learn2sing~\cite{xue2022learn2sing}, the system to teach speakers to sing with a HiFi-GAN vocoder~\cite{kong2020hifi} to synthesize audio from mel-spectrogram. All VITS family systems utilize the same length regulator and duration predictor with our proposed methods instead of Monotonic Alignment Search (MAS). Because the duration of the singing voice is highly related to the music score and MAS will lead to the failure of duration prediction. All the above models are trained with 4 NVIDIA V100 GPUs for fair comparison.


\subsection{Evaluation Metrics}
\vspace{0pt}
We conduct Mean Opinion Score (MOS) and Comparative Mean Opinion Score (CMOS) tests to evaluate the perceptual audio quality and speaker similarity of the synthetic speech and singing samples. The score of MOS test ranges from $1$ to $5$ with an interval of $0.5$, in which $1$ means very bad and $5$ means excellent. The rating score of CMOS ranges from $-3$ to $3$, in which a positive score means the first model is better and vice versa. 
In evaluations, we generate 20 speech and singing samples for each speaker/singer, which are listened to by at least 20 listeners. For an additional subjective evaluation, we calculate the note-level pitch RMSE and Pearson correlation between generated voice and the ground-truth recordings. Note that in the speaker similarity evaluation, the real speech of the two speakers is used as reference for both TTS and SVS tasks as the real singing voices of the speakers are not available. Likewise, the ground truth singing of the two singers are used as reference for both TTS and SVS evaluation.


\section{Results}
\subsection{Subjective Evaluation}


We first conduct MOS evaluation on both TTS and SVS tasks separately in terms of naturalness and speaker similarity, as shown in Table~\ref{tab:mos-tts} and Table~\ref{tab:mos-svs} respectively.

\subsubsection{Evaluation on Text-to-speech.}
According to the MOS results of TTS shown in Table~\ref{tab:mos-tts}, given the target speech data, $\text{VITS}_{\text{TTS}}$, $\text{UniSyn}_{\text{TTS}}$, $\text{VITS}_{\text{unify}}$ and $\text{UniSyn}_{\text{unify}}$ achieve similar scores on both audio naturalness and speaker similarity on Speakers-1 and Speaker-2. It demonstrates that the proposed UniSyn with interpretable latent distribution has the equivalent ability in speech generation with VITS.  
When generating speaking voice for singers, the proposed $\text{UniSyn}_{\text{unify}}$ significantly outperforms $\text{VITS}_{\text{unify}}$ on speech naturalness and speaker similarity. Specifically, the scores of generated speech of singers from $\text{VITS}_{\text{unify}}$ drop significantly compared to the generated speech of speakers, and $\text{UniSyn}_{\text{unify}}$ can still maintain relatively high MOS scores. 
This result shows the superiority of the proposed method in producing synthetic speaking voice of the target singer without speech training data.~\footnote{Audio samples are available at: \url{https://leiyi420.github.io/UniSyn}}

\begin{table}[!htb]
\centering
\small
\renewcommand{\arraystretch}{0.9} 
\setlength{\tabcolsep}{1.2mm}{
\begin{tabular}{l|c|c|c|c}
\toprule
\textbf{}  & \textbf{Speaker-1} & \textbf{Speaker-2} & \textbf{Singer-1}  & \textbf{Singer-2}  \\ 
\midrule
\textbf{}   & \multicolumn{4}{c}{\textbf{Naturalness MOS ($\uparrow$) }}  \\ \midrule
GT  & 4.67±0.08 & 4.62±0.06 & - & -\\ 
\midrule

$\text{VITS}_{\text{ TTS}}$  & 4.17±0.09 & 4.13±0.08 & - & -  \\ 

$\text{VITS}_{\text{ unify}}$ & 4.15±0.09 & 4.14±0.10 & 3.33±0.11 & 3.36±0.12 \\ 
\midrule
$\text{UniSyn}_{\text{ TTS}}$    & 4.18±0.08 & 4.16±0.12 & - & - \\ 

$\text{UniSyn}_{\text{ Unify}}$ & 4.19±0.08 & 4.15±0.09 & \textbf{3.79±0.06} & \textbf{3.81±0.10}\\
\midrule

\textbf{}   & \multicolumn{4}{c}{\textbf{Similarity MOS ($\uparrow$) }}  \\ \toprule

GT & 4.78±0.05 & 4.75±0.07 & - & -  \\ 
\midrule

$\text{VITS}_{\text{TTS}}$  & 4.26±0.13 & 4.25±0.11 & - & -   \\ 
$\text{VITS}_{\text{unify}}$ & 4.23±0.10   & 4.23±0.12 & 3.45±0.11 & 3.48±0.09 \\ 
\midrule

$\text{UniSyn}_{\text{TTS}}$    & 4.27±0.11   & 4.25±0.10 & - & - \\ 

$\text{UniSyn}_{\text{Unify}}$ & 4.24±0.09   & 4.22±0.12 & \textbf{3.71±0.11} & \textbf{3.76±0.12}  \\
\bottomrule

\end{tabular}
}
\caption{Speech naturalness and speaker similarity MOS of TTS with 95\% confidence interval}
\label{tab:mos-tts}
\vspace{-15pt}
\end{table}

\subsubsection{Evaluation on Singing Voice Synthesis.}
We then investigate the performance of singing voice synthesis for all the testing speakers and singers, as shown in Table~\ref{tab:mos-svs}. For the singer singing task, the UniSyn family can be on par with the VITS family, and they are much better than the baseline Learn2Sing. For the speaker singing task, the naturalness scores of $\text{VITS}_{\text{unify}}$ and Leanr2Sing are much lower than the singer singing task, where the produced singing voice of the speakers has obvious artifacts. $\text{UniSyn}_{\text{unify}}$ surpasses VITS and Leanr2Sing by a large margin when generating speaker's singing voice. As for speaker similarity, the scores of Learn2Sing, VITS, and Unisyn are close, which means the three models have similar ability to maintain speaker timbre. 
The results of naturalness and speaker similarity indicate that the UniSyn can generate natural singing voices for the speakers, even if they do not have any singing training data. It is worth noticing that the speaker similarity scores of speakers are much lower than the singers in the SVS task. This is mainly because there is no ground-truth singing voice for the speakers, he listeners only can judge the similarity with the speaking data, which has obvious differences from their singing speech.



\begin{table}[!htb]
\centering
\small
\renewcommand{\arraystretch}{0.9} 
\setlength{\tabcolsep}{1.3mm}{
\begin{tabular}{l|c|c|c|c}
\toprule
\textbf{}  & \textbf{Speaker-1} & \textbf{Speaker-2} & \textbf{Singer-1}  & \textbf{Singer-2} \\ 
\midrule
\textbf{}   & \multicolumn{4}{c}{\textbf{Naturalness MOS ($\uparrow$) }} \\ \midrule

GT   & - & - & 4.65±0.08 & 4.78±0.05\\ 
\midrule

\text{Learn2Sing}   & 3.07±0.16 & 3.03±0.15 & 3.85±0.08 & 3.87±0.09 \\ 
\midrule


$\text{VITS}_\text{SVS}$   & - & - & 3.95±0.08 & 3.97±0.09\\ 

$\text{VITS}_\text{unify}$  & 3.15±0.10 & 3.05±0.09 & 3.93±0.11 & 3.95±0.12 \\ 
\midrule

$\text{UniSyn}_\text{SVS}$    & - & - & 3.95±0.06 & 3.96±0.07\\ 

$\text{UniSyn}_\text{unify}$   & \textbf{3.73±0.11} & \textbf{3.78±0.10} & 3.95±0.06 & 3.97±0.08  \\
\midrule
\textbf{}   & \multicolumn{4}{c}{\textbf{Speaker similarity MOS ($\uparrow$) }} \\ \midrule

GT  & - & - & 4.57±0.06 & 4.61±0.04\\ 
\midrule

\text{Learn2Sing} & 3.75±0.12 & 3.76±0.15 & 3.89±0.12 & 3.93±0.08  \\ 
\midrule


$\text{VITS}_\text{SVS}$ & - & - & 4.15±0.08 & 4.17±0.07  \\ 

$\text{VITS}_\text{unify}$  & 3.75±0.11   & \textbf{3.79±0.13} & 4.15±0.13 & 4.16±0.11  \\ 
\midrule

$\text{UniSyn}_\text{SVS}$   & -   & - & 4.17±0.07 & 4.18±0.09 \\ 

$\text{UniSyn}_\text{unify}$  & \textbf{3.76±0.10}   & 3.78±0.12 & 4.14±0.08 & 4.18±0.07  \\
\bottomrule

\end{tabular}
}
\caption{Speech naturalness and speaker similarity MOS of SVS with 95\% confidence interval.}
\label{tab:mos-svs}
\vspace{-15pt}
\end{table}




\subsection{Objective Evaluation}
Since pitch is critical to the singing voice, we further conduct objective evaluations to measure the pitch accuracy of the synthesized singing voice, as shown in Table~\ref{tab:f0}. The measurements include root mean squared error (RMSE) and Pearson correlation (Corr) with the ground-truth recordings. For Singer-1 and Singer-2, all the models achieve a similar level of RMSE and Corr on pitch. While for the singing voice of Speaker-1 and Speaker-2, the proposed $\text{UniSyn}_{\text{unify}}$ achieves the lowest RMSE and highest Corr results, which reaches the same level as those of the singer's singing voice. But for Learn2Sing and VITS, RMSE remains very high while Corr is obviously at a lower level. The objective evaluation results are consistent with the subjective tests and further prove that our unified model can synthesize natural singing voices for the speakers without singing training data.

\begin{table}[!htb]\centering
\centering
\small
\renewcommand{\arraystretch}{0.8} 
\setlength{\tabcolsep}{1.5 mm}{
\begin{tabular}{l|cccc}
\toprule
 & \textbf{Speaker-1} & \textbf{Speaker-2} & \textbf{Singer-1} & \textbf{Singer-2}  \\ 
\midrule
\multicolumn{5}{l}{\textbf{\text{Learn2Sing}}}  \\
\midrule
 RMSE ($\downarrow$) & 29.559 & 33.554 & 10.868 & 10.280 \\
 Corr ($\uparrow$) & 0.837 & 0.813 & 0.922 & 0.925 \\ 
\midrule
\multicolumn{3}{l}{\textbf{$\text{VITS}_\text{unify}$}} \\
\midrule
 RMSE ($\downarrow$) & 24.651 & 29.458 & 9.012 & 9.389 \\
 Corr ($\uparrow$) & 0.866 & 0.835 & 0.964 & 0.963 \\ 
\midrule
\multicolumn{3}{l}{\textbf{$\text{UniSyn}_\text{unify}$}}  \\
\midrule
 RMSE ($\downarrow$) & \textbf{8.672} & \textbf{9.954} & \textbf{8.698} & \textbf{8.738} \\
 Corr ($\uparrow$) & \textbf{0.967} & \textbf{0.957} & \textbf{0.967} & \textbf{0.967} \\ 
\bottomrule
\end{tabular}
}
\caption{Objective evaluation on pitch, where ``RMSE" and ``Corr" denote the RMSE and Pearson correlation of pitch.}
\label{tab:f0}
\vspace{-15pt}
\end{table}

\subsection{Ablation Studies}
The CMOS results of ablation studies are illustrated in Table~\ref{tab:abation}, where ``-pert" and ``-GVAE" denote removing speaker perturbation and supervised guided-VAE from UniSyn respectively. A more negative CMOS score means UniSyn is much better.
From the TTS of speakers and SVS of singers results, we find removing the perturbation or GVAE would not have much impact on the synthetic voices, which indicates the proposed MC-VAE can effectively conduct voice generation. 
From the cross-over TTS and SVS, it can be seen that: 1) removing the speaker perturbation strategy leads to a significant decline in speaker similarity, which indicates that speaker perturbation mainly contributes to the timbre during cross-speaker TTS; 2) removing GVAE results in a significant drop on naturalness and also noticeable influence on speaker similarity, which demonstrates that GVAE plays an important role in keeping naturalness and speaker similarity; 3) removing both components have medium scores compared with individually removing them.

We also notice that removing the two strategies leads to different degrees of performance degradation. To be specific, removing speaker timbre perturbation 
noticeably leads to different degree of degradation on speaker similarity for TTS and SVS. We also find that GVAE is more important to the naturalness of the TTS task although it is necessary for both tasks and metrics.




\begin{table}[!htb]\centering
\centering
\small
\renewcommand{\arraystretch}{0.8} 
 \setlength{\tabcolsep}{1.8mm}{
\begin{tabular}{l|c|cc}
\toprule
\multicolumn{2}{c}{}      & \textbf{Naturalness ($\uparrow$) } & \textbf{Similarity ($\uparrow$) } \\
\midrule
\multicolumn{4}{l}{\textbf{TTS}} \\
\midrule
\multirow{2}{*}{-pert} & Speakers  & -0.006 & -0.012  \\
& Singers  & 0.002   & -0.356  \\ \midrule

\multirow{2}{*}{-GVAE} & Speakers & -0.015 & -0.023 \\
 & Singers  & -0.430  &  -0.164  \\ \midrule
 
\multirow{2}{*}{-pert-GVAE}  & Speakers  & -0.015     & -0.014 \\
& Singers  & -0.347  &  -0.224   \\\midrule

\multicolumn{4}{l}{\textbf{SVS}}  \\
\midrule
\multirow{2}{*}{-pert} & Speakers & -0.004 & -0.235 \\
& Singers & -0.021 & -0.018      \\
\midrule
\multirow{2}{*}{-GVAE} & Speakers & -0.163 & -0.102 \\
& Singers & -0.012 & -0.007     \\
\midrule
\multirow{2}{*}{-pert-GVAE}         & Speskers  & -0.104        & -0.106     \\
 & Singers  & -0.021       &  -0.036          \\

\bottomrule
\end{tabular}
}
\caption{CMOS values for ablation studies of UniSyn}
\label{tab:abation}
\end{table}




\vspace{-15pt}
\section{Conclusions}
In this work, we propose \textit{UniSyn} to conduct TTS and SVS in an end-to-end unified trained neural model. UniSyn also has the advantage of synthesizing both speaking and singing voice with only speaking or only singing training data of a target person. Based on a specifically designed unified textual representation, we propose MC-VAE to create a more interpretable latent space for speaking/singing style transfer across speakers. We further introduce GVAE and timbre perturbation into UniSyn to improve the speaker similarity and naturalness of synthetic speech and singing. Experiments and ablation studies show that UniSyn outperforms the state-of-the-art end-to-end synthesis framework, VITS, on both speaker singing and singer speaking tasks, which indicates the effectiveness of the proposed approach in the unified modeling of TTS and SVS.

\section{Acknowledgment}
This work was supported by the National Key Research and Development Program of China (No. 2020AAA0108600). Thanks to participants in the listening test for the valuable evaluations.

\bibliography{mybibfile}

\begin{thebibliography}{40}
\providecommand{\natexlab}[1]{#1}

\bibitem[{Agarwal et~al.(2019)Agarwal, Liang, Schuurmans, and
  Norouzi}]{agarwal2019learning}
Agarwal, R.; Liang, C.; Schuurmans, D.; and Norouzi, M. 2019.
\newblock Learning to generalize from sparse and underspecified rewards.
\newblock In \emph{Proceedings of International conference on machine
  learning}, 130--140. PMLR.

\bibitem[{Blaauw and Bonada(2020)}]{blaauw2020sequence}
Blaauw, M.; and Bonada, J. 2020.
\newblock Sequence-to-sequence singing synthesis using the feed-forward
  transformer.
\newblock In \emph{Proceedings of International Conference on Acoustics, Speech
  and Signal Processing (ICASSP)}, 7229--7233. IEEE.

\bibitem[{Boersma and Van~Heuven(2001)}]{boersma2001speak}
Boersma, P.; and Van~Heuven, V. 2001.
\newblock Speak and unSpeak with PRAAT.
\newblock \emph{Glot International}, 5(9/10): 341--347.

\bibitem[{Chen et~al.(2021)Chen, Tan, Li, Liu, Qin, Zhao, and
  Liu}]{chen2021adaspeech}
Chen, M.; Tan, X.; Li, B.; Liu, Y.; Qin, T.; Zhao, S.; and Liu, T.-Y. 2021.
\newblock Adaspeech: Adaptive text to speech for custom voice.
\newblock arXiv:2203.16408.

\bibitem[{Choi et~al.(2021)Choi, Lee, Kim, Lee, Heo, and Lee}]{choi2021neural}
Choi, H.-S.; Lee, J.; Kim, W.; Lee, J.; Heo, H.; and Lee, K. 2021.
\newblock Neural analysis and synthesis: Reconstructing speech from
  self-supervised representations.
\newblock \emph{Advances in Neural Information Processing Systems}, 34:
  16251--16265.

\bibitem[{Ding et~al.(2020)Ding, Xu, Xu, Parmar, Yang, Welling, and
  Tu}]{ding2020guided}
Ding, Z.; Xu, Y.; Xu, W.; Parmar, G.; Yang, Y.; Welling, M.; and Tu, Z. 2020.
\newblock Guided variational autoencoder for disentanglement learning.
\newblock In \emph{Proceedings of the IEEE/CVF Conference on Computer Vision
  and Pattern Recognition}, 7920--7929.

\bibitem[{Gu et~al.(2021)Gu, Yin, Rao, Wan, Tang, Zhang, Chen, Wang, and
  Ma}]{gu2021bytesing}
Gu, Y.; Yin, X.; Rao, Y.; Wan, Y.; Tang, B.; Zhang, Y.; Chen, J.; Wang, Y.; and
  Ma, Z. 2021.
\newblock Bytesing: A chinese singing voice synthesis system using duration
  allocated encoder-decoder acoustic models and wavernn vocoders.
\newblock In \emph{Proceedings of International Symposium on Chinese Spoken
  Language Processing (ISCSLP)}, 1--5. IEEE.

\bibitem[{Guo et~al.(2022)Guo, Zhou, Meng, and Liu}]{guo2022improving}
Guo, H.; Zhou, Z.; Meng, F.; and Liu, K. 2022.
\newblock Improving Adversarial Waveform Generation Based Singing Voice
  Conversion with Harmonic Signals.
\newblock In \emph{Proceedings of International Conference on Acoustics, Speech
  and Signal Processing (ICASSP)}, 6657--6661. IEEE.

\bibitem[{Jeong et~al.(2021)Jeong, Kim, Cheon, Choi, and Kim}]{jeong2021diff}
Jeong, M.; Kim, H.; Cheon, S.~J.; Choi, B.~J.; and Kim, N.~S. 2021.
\newblock Diff-tts: A denoising diffusion model for text-to-speech.
\newblock arXiv:2104.01409.

\bibitem[{Kim et~al.(2018)Kim, Choi, Park, Hahn, Kim, and Kim}]{kim2018korean}
Kim, J.; Choi, H.; Park, J.; Hahn, M.; Kim, S.; and Kim, J.-J. 2018.
\newblock Korean singing voice synthesis system based on an LSTM recurrent
  neural network.
\newblock In \emph{Proceedings of INTERSPEECH}, 1551--1555.

\bibitem[{Kim, Kong, and Son(2021)}]{kim2021conditional}
Kim, J.; Kong, J.; and Son, J. 2021.
\newblock Conditional variational autoencoder with adversarial learning for
  end-to-end text-to-speech.
\newblock In \emph{Proceedings of International Conference on Machine
  Learning}, 5530--5540. PMLR.

\bibitem[{Kong, Kim, and Bae(2020)}]{kong2020hifi}
Kong, J.; Kim, J.; and Bae, J. 2020.
\newblock HiFi-GAN: Generative adversarial networks for efficient and high
  fidelity speech synthesis.
\newblock 17022--17033.

\bibitem[{Lee et~al.(2020)Lee, Choi, Koo, and Lee}]{lee2020disentangling}
Lee, J.; Choi, H.-S.; Koo, J.; and Lee, K. 2020.
\newblock Disentangling timbre and singing style with multi-singer singing
  synthesis system.
\newblock In \emph{the International Conference on Acoustics, Speech, and
  Signal Processing (ICASSP)}, 7224--7228. IEEE.

\bibitem[{Li et~al.(2019)Li, Liu, Liu, Zhao, and Liu}]{li2019neural}
Li, N.; Liu, S.; Liu, Y.; Zhao, S.; and Liu, M. 2019.
\newblock Neural speech synthesis with transformer network.
\newblock In \emph{Proceedings of the AAAI Conference on Artificial
  Intelligence}, volume~33, 6706--6713.

\bibitem[{Li et~al.(2021)Li, Tang, Yin, Wan, Xu, Shen, and Ma}]{li2021ppg}
Li, Z.; Tang, B.; Yin, X.; Wan, Y.; Xu, L.; Shen, C.; and Ma, Z. 2021.
\newblock Ppg-based singing voice conversion with adversarial representation
  learning.
\newblock In \emph{Proceedings of International Conference on Acoustics, Speech
  and Signal Processing (ICASSP)}, 7073--7077. IEEE.

\bibitem[{Liu et~al.(2022)Liu, Li, Ren, Chen, and Zhao}]{liu2022diffsinger}
Liu, J.; Li, C.; Ren, Y.; Chen, F.; and Zhao, Z. 2022.
\newblock Diffsinger: Singing voice synthesis via shallow diffusion mechanism.
\newblock In \emph{Proceedings of the AAAI Conference on Artificial
  Intelligence}, volume~36, 11020--11028.

\bibitem[{Liu et~al.(2021{\natexlab{a}})Liu, Cao, Liu, Hu, Li, Weng, and
  Su}]{liu2021vara}
Liu, P.; Cao, Y.; Liu, S.; Hu, N.; Li, G.; Weng, C.; and Su, D.
  2021{\natexlab{a}}.
\newblock Vara-tts: Non-autoregressive text-to-speech synthesis based on very
  deep vae with residual attention.
\newblock arXiv:2102.06431.

\bibitem[{Liu et~al.(2021{\natexlab{b}})Liu, Wen, Lu, Song, and
  Sung}]{liu2021vibrato}
Liu, R.; Wen, X.; Lu, C.; Song, L.; and Sung, J.~S. 2021{\natexlab{b}}.
\newblock Vibrato Learning in Multi-Singer Singing Voice Synthesis.
\newblock In \emph{Proceedings of Automatic Speech Recognition and
  Understanding Workshop (ASRU)}, 773--779. IEEE.

\bibitem[{Liu et~al.(2021{\natexlab{c}})Liu, Cao, Hu, Su, and
  Meng}]{liu2021fastsvc}
Liu, S.; Cao, Y.; Hu, N.; Su, D.; and Meng, H. 2021{\natexlab{c}}.
\newblock Fastsvc: Fast cross-domain singing voice conversion with feature-wise
  linear modulation.
\newblock In \emph{Proceedings of International Conference on Multimedia and
  Expo (ICME)}, 1--6. IEEE.

\bibitem[{Lu et~al.(2020)Lu, Wu, Luan, Tan, and Zhou}]{lu2020xiaoicesing}
Lu, P.; Wu, J.; Luan, J.; Tan, X.; and Zhou, L. 2020.
\newblock Xiaoicesing: A high-quality and integrated singing voice synthesis
  system.
\newblock arXiv:2006.06261.

\bibitem[{Luo et~al.(2020)Luo, Hsu, Agres, and Herremans}]{luo2020singing}
Luo, Y.-J.; Hsu, C.-C.; Agres, K.; and Herremans, D. 2020.
\newblock Singing voice conversion with disentangled representations of singer
  and vocal technique using variational autoencoders.
\newblock In \emph{Proceedings of International Conference on Acoustics, Speech
  and Signal Processing (ICASSP)}, 3277--3281. IEEE.

\bibitem[{Mei et~al.(2019)Mei, Xiao, Huang, Schuurmans, and
  M{\"u}ller}]{mei2019principled}
Mei, J.; Xiao, C.; Huang, R.; Schuurmans, D.; and M{\"u}ller, M. 2019.
\newblock On principled entropy exploration in policy optimization.
\newblock In \emph{Proceedings of International Joint Conference on Artificial
  Intelligence}, 3130--3136.

\bibitem[{Morise, Yokomori, and Ozawa(2016)}]{morise2016world}
Morise, M.; Yokomori, F.; and Ozawa, K. 2016.
\newblock WORLD: a vocoder-based high-quality speech synthesis system for
  real-time applications.
\newblock \emph{IEICE TRANSACTIONS on Information and Systems}, 99(7):
  1877--1884.

\bibitem[{Nishimura et~al.(2016)Nishimura, Hashimoto, Oura, Nankaku, and
  Tokuda}]{nishimura2016singing}
Nishimura, M.; Hashimoto, K.; Oura, K.; Nankaku, Y.; and Tokuda, K. 2016.
\newblock Singing Voice Synthesis Based on Deep Neural Networks.
\newblock In \emph{Interspeech}, 2478--2482.

\bibitem[{Polyak et~al.(2020)Polyak, Wolf, Adi, and Taigman}]{Polyak2020}
Polyak, A.; Wolf, L.; Adi, Y.; and Taigman, Y. 2020.
\newblock {Unsupervised Cross-Domain Singing Voice Conversion}.
\newblock In \emph{Proceedings of INTERSPEECH}, 801--805.

\bibitem[{Ren et~al.(2019)Ren, Ruan, Tan, Qin, Zhao, Zhao, and
  Liu}]{ren2019fastspeech}
Ren, Y.; Ruan, Y.; Tan, X.; Qin, T.; Zhao, S.; Zhao, Z.; and Liu, T.-Y. 2019.
\newblock Fastspeech: Fast, robust and controllable text to speech.
\newblock 3165--3174.

\bibitem[{Ren et~al.(2020)Ren, Tan, Qin, Luan, Zhao, and
  Liu}]{ren2020deepsinger}
Ren, Y.; Tan, X.; Qin, T.; Luan, J.; Zhao, Z.; and Liu, T.-Y. 2020.
\newblock Deepsinger: Singing voice synthesis with data mined from the web.
\newblock In \emph{Proceedings of ACM SIGKDD International Conference on
  Knowledge Discovery \& Data Mining}, 1979--1989.

\bibitem[{Saito et~al.(2018)Saito, Ijima, Nishida, and
  Takamichi}]{saito2018non}
Saito, Y.; Ijima, Y.; Nishida, K.; and Takamichi, S. 2018.
\newblock Non-parallel voice conversion using variational autoencoders
  conditioned by phonetic posteriorgrams and d-vectors.
\newblock In \emph{Proceedings of International Conference on Acoustics, Speech
  and Signal Processing (ICASSP)}, 5274--5278. IEEE.

\bibitem[{Shen et~al.(2018)Shen, Pang, Weiss, Schuster, Jaitly, Yang, Chen,
  Zhang, Wang, Skerrv-Ryan et~al.}]{shen2018natural}
Shen, J.; Pang, R.; Weiss, R.~J.; Schuster, M.; Jaitly, N.; Yang, Z.; Chen, Z.;
  Zhang, Y.; Wang, Y.; Skerrv-Ryan, R.; et~al. 2018.
\newblock Natural tts synthesis by conditioning wavenet on mel spectrogram
  predictions.
\newblock In \emph{Proceedings of International Conference on Acoustics, Speech
  and Signal Processing (ICASSP)}, 4779--4783. IEEE.

\bibitem[{Sj{\"o}lander(2003)}]{sjolander2003hmm}
Sj{\"o}lander, K. 2003.
\newblock An HMM-based system for automatic segmentation and alignment of
  speech.
\newblock In \emph{Proceedings of fonetik}, volume 2003, 93--96. Citeseer.

\bibitem[{Valle et~al.(2020)Valle, Li, Prenger, and
  Catanzaro}]{valle2020mellotron}
Valle, R.; Li, J.; Prenger, R.; and Catanzaro, B. 2020.
\newblock Mellotron: Multispeaker expressive voice synthesis by conditioning on
  rhythm, pitch and global style tokens.
\newblock In \emph{Proceedings of International Conference on Acoustics, Speech
  and Signal Processing (ICASSP)}, 6189--6193. IEEE.

\bibitem[{Vaswani et~al.(2022)Vaswani, Bachem, Totaro, M{\"u}ller, Garg, Geist,
  Machado, Castro, and Le~Roux}]{vaswani2022general}
Vaswani, S.; Bachem, O.; Totaro, S.; M{\"u}ller, R.; Garg, S.; Geist, M.;
  Machado, M.~C.; Castro, P.~S.; and Le~Roux, N. 2022.
\newblock A general class of surrogate functions for stable and efficient
  reinforcement learning.
\newblock In \emph{Proceedings of AISTATS}, 8619--8649.

\bibitem[{Wang et~al.(2017)Wang, Skerry-Ryan, Stanton, Wu, Weiss, Jaitly, Yang
  et~al.}]{wang2017tacotron}
Wang, Y.; Skerry-Ryan, R.; Stanton, D.; Wu, Y.; Weiss, R.~J.; Jaitly, N.; Yang,
  Z.; et~al. 2017.
\newblock {Tacotron: Towards end-to-end speech synthesis}.
\newblock In \emph{Proceedings of INTERSPEECH}, 4006--4010.

\bibitem[{Wang et~al.(2022)Wang, Wang, Zhu, Wu, Li, Xue, Zhang, Xie, and
  Bi}]{wang2022opencpop}
Wang, Y.; Wang, X.; Zhu, P.; Wu, J.; Li, H.; Xue, H.; Zhang, Y.; Xie, L.; and
  Bi, M. 2022.
\newblock Opencpop: A High-Quality Open Source Chinese Popular Song Corpus for
  Singing Voice Synthesis.
\newblock In \emph{Proc. Interspeech 2022}, 4242--4246.

\bibitem[{Weiss et~al.(2021)Weiss, Skerry-Ryan, Battenberg, Mariooryad, and
  Kingma}]{weiss2021wave}
Weiss, R.~J.; Skerry-Ryan, R.; Battenberg, E.; Mariooryad, S.; and Kingma,
  D.~P. 2021.
\newblock Wave-tacotron: Spectrogram-free end-to-end text-to-speech synthesis.
\newblock In \emph{Proceedings of International Conference on Acoustics, Speech
  and Signal Processing (ICASSP)}, 5679--5683. IEEE.

\bibitem[{Xue et~al.(2022)Xue, Wang, Zhang, Xie, Zhu, and
  Bi}]{xue2022learn2sing}
Xue, H.; Wang, X.; Zhang, Y.; Xie, L.; Zhu, P.; and Bi, M. 2022.
\newblock {Learn2Sing 2.0: Diffusion and Mutual Information-Based Target
  Speaker SVS by Learning from Singing Teacher}.
\newblock In \emph{Proc. Interspeech 2022}, 4267--4271.

\bibitem[{Xue et~al.(2021)Xue, Yang, Lei, Xie, and Li}]{xue2021learn2sing}
Xue, H.; Yang, S.; Lei, Y.; Xie, L.; and Li, X. 2021.
\newblock Learn2sing: Target speaker singing voice synthesis by learning from a
  singing teacher.
\newblock In \emph{Proceedings of IEEE Spoken Language Technology Workshop
  (SLT)}, 522--529. IEEE.

\bibitem[{Zhang et~al.(2020)Zhang, Yu, Lu, Weng, Zhang, Wu, Xie, Li, and
  Yu}]{Zhang2020}
Zhang, L.; Yu, C.; Lu, H.; Weng, C.; Zhang, C.; Wu, Y.; Xie, X.; Li, Z.; and
  Yu, D. 2020.
\newblock {DurIAN-SC: Duration Informed Attention Network Based Singing Voice
  Conversion System}.
\newblock In \emph{Proceedings of INTERSPEECH}, 1231--1235.

\bibitem[{Zhang et~al.(2022)Zhang, Cong, Xue, Xie, Zhu, and
  Bi}]{zhang2022visinger}
Zhang, Y.; Cong, J.; Xue, H.; Xie, L.; Zhu, P.; and Bi, M. 2022.
\newblock Visinger: Variational inference with adversarial learning for
  end-to-end singing voice synthesis.
\newblock In \emph{Proceedings of International Conference on Acoustics, Speech
  and Signal Processing (ICASSP)}, 7237--7241. IEEE.

\bibitem[{Zhao et~al.(2018)Zhao, Kim, Zhang, Rush, and
  LeCun}]{zhao2018adversarially}
Zhao, J.; Kim, Y.; Zhang, K.; Rush, A.; and LeCun, Y. 2018.
\newblock Adversarially regularized autoencoders.
\newblock In \emph{Proceedings of International conference on machine
  learning}, 5902--5911. PMLR.

\end{thebibliography}

\appendix

\end{document}